\documentclass[conference]{IEEEtran}
\IEEEoverridecommandlockouts

\usepackage{amssymb}
\usepackage[cmex10]{amsmath}
\usepackage{stfloats}
\usepackage{graphicx}
\usepackage{subfigure}
\usepackage{tabularx}
\usepackage{epsfig,epsf,color,balance,cite}
\usepackage{verbatim}
\usepackage{url}
\usepackage{bm}
\usepackage{booktabs}
\usepackage[ruled,vlined]{algorithm2e}
\usepackage{algorithmic}
\usepackage{geometry}
\usepackage{color}
\definecolor{myc1}{rgb}{0,0,0}

\begin{document}

% paper title
\title{Fair Rate Maximization for Fluid Antenna Relay (FAR)-assisted Multi-user MISO Communications}

\author{
\IEEEauthorblockN{
Ruopeng Xu$\IEEEauthorrefmark{1}$$\IEEEauthorrefmark{2}$,
Zhaohui Yang$\IEEEauthorrefmark{1}$$\IEEEauthorrefmark{2}$,
Ting Zhang $\IEEEauthorrefmark{1}$$\IEEEauthorrefmark{2}$,
Mingzhe Chen$\IEEEauthorrefmark{3}$,
Chen Zhu$\IEEEauthorrefmark{4}$,
Zhaoyang Zhang$\IEEEauthorrefmark{1}$$\IEEEauthorrefmark{2}$}
	\IEEEauthorblockA{
			$\IEEEauthorrefmark{1}$College of Information Science and Electronic Engineering, Zhejiang University, Hangzhou, China\\
			$\IEEEauthorrefmark{2}$Zhejiang Provincial Key Laboratory of Info. Proc., Commun. \& Netw. (IPCAN), Hangzhou, China\\
            $\IEEEauthorrefmark{3}$Department of Electrical and Computer Engineering and Institute for Data Science and Computing, University of Miami\\
            $\IEEEauthorrefmark{4}$Polytechnic Institute, Zhejiang University, Hangzhou, China\\
  % $\IEEEauthorrefmark{5}$Zhejiang Lab, Hangzhou, China\\
          	E-mails:
ruopengxu@zju.edu.cn,
yang\_zhaohui@zju.edu.cn,
zhang\_ting@zju.edu.cn,\\
mingzhe.chen@miami.edu,
zhuc@zju.edu.cn,
ning\_ming@zju.edu.cn
 %dr.h.chen@ieee.org
		}
% \thanks{This work is supported by Zhejiang Lab Program under grant K2023QA0AL02, and Zhejiang Science and Technology Program under grant 2023C01021.}
\vspace{-2em}
}
%Hongyang Chen$\IEEEauthorrefmark{5}$, 
% make the title area
\maketitle

\begin{abstract}
In this paper, we investigate the problem of max-min rate maximization in fluid antenna relay (FAR)-assisted multi-user uplink multiple-input single-output (MISO) wireless systems, where each user is equipped with a single fluid antenna (FA) and the base station (BS) is equipped with multiple FAs. Unlike most existing relevant work focusing on maximizing sum rate of the fluid antenna system (FAS), which may cause unbearable rate loss to weak users, we propose to maximize the minimal rate of the system to ensure fairness. The max-min optimization problem is formulated by jointly optimizing the positions of FAs with meeting the minimum distance requirements of FAs, maximum transmitting power limit, and feasible antenna region constraints. To solve this problem, we propose an alternating algorithm with utilizing the successive convex approximation (SCA) method. Simulation results demonstrate that the proposed method significantly outperforms conventional methods in terms of maximizing the minimal achievable rate across different signal-to-noise ratios (SNRs) and normalized region sizes.
\end{abstract}

\IEEEpeerreviewmaketitle

\section{Introduction}
With the rapid development of the sixth generation (6G) wireless communication technology in various fields\cite{xu2023edge,yang2023energy}, the fluid antenna system (FAS) has received widespread attention. Unlike traditional antennas, fluid antennas (FAs) can freely and instantly switch their positions within a given region. The work in \cite{wong2020fluid} first proposes FAS that contains a single antenna enabling changing its position in a small linear space. Soon afterward, many research works revolved around FAS. While research on flexible FAS architectures, such as liquid-based antennas\cite{abu2021liquid} and pixel-based antennas\cite{zhang2024pixel}, enables the possibility of FAs' instant position change, the work in \cite{wong2020performance,khammassi2023new,chai2022port} promotes the development of theoretical research in FAS. Authors in \cite{wong2020performance} prove that the FAS outperforms traditional multiple antenna systems even with a tiny region by deriving the ergodic capacity and a capacity lower bound. The study in \cite{khammassi2023new} provides a superior two-stage analytical approximation of the FAS channel. The work in \cite{chai2022port} investigates the best port selection of antenna position.

The development of FAS and other fields is constantly promoting each other. FAS empowers the development of multiple-input multiple-output (MIMO) system\cite{wang2024fluid}, integrated sensing and communication (ISAC) system\cite{zhou2024fluid}, and near field system\cite{chen2024joint}. The work in \cite{wang2024fluid} jointly optimizes antenna ports at the FAS with the precoding design to achieve a higher sum rate. The study in \cite{zhou2024fluid} maximizes the achievable rate by jointly optimizing the transmitting beamforming and FA locations. Authors in \cite{chen2024joint} investigate the energy efficiency maximization problem for FAS in near field communications. Meanwhile, the development of other regions, such as artificial intelligence (AI)\cite{wang2024ai}, has also driven the research on FAS.

However, existing research has focused mainly on maximizing the sum rate of the system\cite{zhou2024fluid,ye2023fluid}, which may result in intolerable rate loss for users with poor channel conditions in multi-user scenarios. To ensure the achievable rate of the weak user, we propose a max-min optimization to maximize the minimal rate of the system. Besides, existing research typically assumes a line-of-sight (LoS) link between the transmitter and the receiver, neglecting the inevitable obstacles that may cause only non-LoS (NLoS) links to exist during transmission. The work in \cite{10615841} makes the first attempt to deploy FAs in the blockage as a relay to maximize the system sum rate, but also did not consider the fairness of users with poor channel conditions. Thus, we propose to consider the fair uplink wireless communication assisted by the fluid antenna relay (FAR).

The main contributions of this paper are listed as follows:
\begin{itemize}
    \item We investigate an uplink FAR-assisted multi-user multiple-input single-output (MISO) communication system, where FAR is introduced as a relay to improve signal transmission in response to the issues caused by LoS path blockage. In the considered model, users are equipped with single FA while FAR and base station (BS) are deployed multiple FAs. We model the channel during the transmission and then formulate a max-min optimization problem to maximize the minimal rate of the system through jointly optimizing positions of FAs.
    \item To solve this max-min fairness problem, we first transform it into a problem of maximizing channel gain. To address this maximization issue, an alternating optimization algorithm is proposed through iteratively solving a sequence of sub-problems with the successive convex approximation (SCA).
    \item Simulation results demonstrate the effectiveness of the proposed algorithm in enhancing the minimal achievable rate. Compared to conventional schemes across, the proposed algorithm always outperforms other baselines no matter signal-to-noise ratio (SNR) or normalized region size changes, showcasing its robustness and superiority.
\end{itemize}
\textit{Notation}: We use lower case letters to denote scalars, bold lower case letters to denote vectors, and bold upper letters to denote matrices. The subscripts $(\cdot)^{T}$ and $(\cdot)^{H}$ are denoted as the transpose and conjugate transpose (Hermitian) operations, respectively. $\bm{I}_N$ denotes an $N \times N$ dimensional identity matrix, $\mathcal{M}$ denotes the number set $\{1,\dots,M\}$, and the sets of $M \times N$ dimensional complex and real matrices are denoted by $\mathbb{C}^{M \times N}$ and $\mathbb{R}^{M \times N}$, respectively. The operation 
$\mathrm{diag}(\bm{a})$ generates a diagonal matrix with the elements of $\bm{a}$ along its main diagonal, the operation $\mathrm{Re}(a)$ obtains the real part of scaler $a$, and the operation $\mathrm{tr}(\bm{A})$ generates the trace of $\bm{A}$. $\mathcal{CN}(\bm{a},\bm{B})$ represents the symmetric complex-valued Gaussian distribution with mean $\bm{a}$ and covariance matrix $\bm{B}$. $||\bm{a}||_2$ stands for the Euclidean norm of $\bm{a}$.

\section{System Model}
\subsection{FAR-assisted MISO System}
\begin{figure}[t]
\centering
\includegraphics[width=1\linewidth]{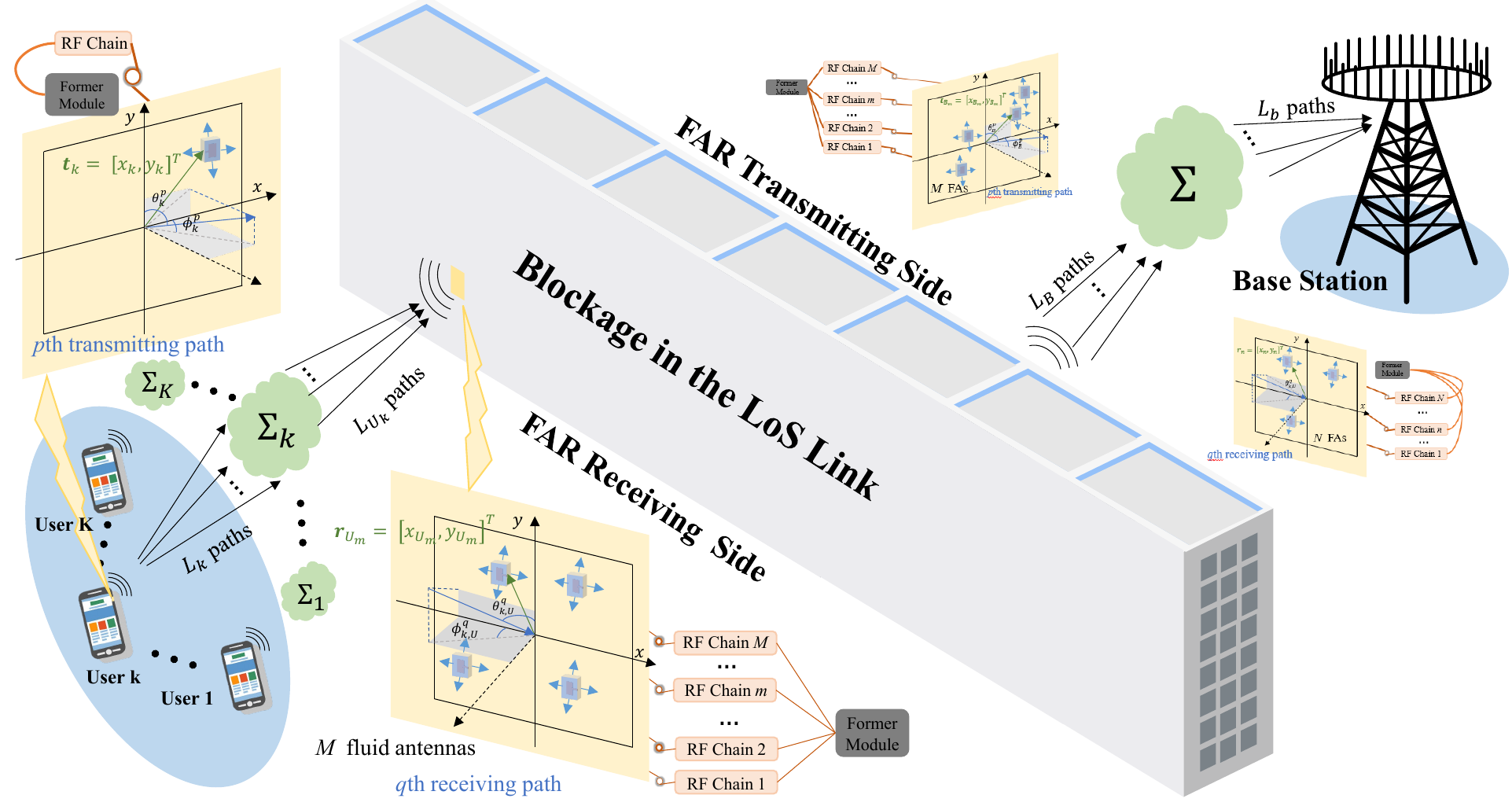}
\caption{FAR-assisted wireless communication system} 
\label{System Model}
\end{figure}

As depicted in Fig.~\ref{System Model}, we consider an FAR-assisted wireless communication system, comprising $K$ single-FA users, an FAR equipped with $M$ FAs at both sides, and a BS with $N$ FAs. Due to the blockage in the LoS path, we employ $M$ FAs at one side of the blockage to receive the signals from the users (receiving side, RS), and $M$ FAs at the other side (transmitting side, TS) to transmit the signals to the BS. We assume that FAs of the users, FAR, and BS are all connected to radio frequency chains via flexible cables, so that they can freely adjust their positions at a given region\cite{ma2023mimo}, which is of size $A \times A$. To clearly represent the positions of FAs, we introduce the position vector in two-dimensional Cartesian coordinate to locate the relative position of each FA. For example, the position vector of the FA of user $k$ is $\bm{t}_k = [x_k,y_k]^T\in \mathbb{R}^{2 \times 1}$. 
%\addtolength{\topmargin}{-0.08in}
\subsection{Channel Model}
Consider the two-dimensional plane where FA can move freely. The elevation and azimuth angle of departure (AoD) can be denoted as $\theta_{k}^p \in [0,\pi]$ and $\phi_{k}^p \in [0,\pi]$, where the subscript $k$ denotes user $k$ and superscript $p$ denotes the $p$th transmitting path. For the convenience of representation, we introduce wave vector $\bm{n}_k^p = [\mathrm{sin}\theta_{k}^p\mathrm{cos}\phi_{k}^p,\mathrm{cos}\theta_{k}^p]^T$ to represent the $p$th transmitting path of user $k$, and the phase difference of the signal propagation for the $p$th transmitting path is
\begin{equation}\label{phaseDifference}
    \rho_k^p(\bm{t}_k) = e^{j \frac{2\pi}{\lambda}(\bm{n}_k^p)^T \bm{t}_k},
\end{equation}
where $\lambda $ is the carrier wavelength. 

Then, the field response vector (FRV) of all the transmitting paths of user $k$ can be defined as 
\begin{equation}
    \bm{u}_k (\bm{t}_k)  = [\rho_k^1 (\bm{t}_k), \rho_k^2 (\bm{t}_k), \dots, \rho_k^{L_k} (\bm{t}_k)]^T \in \mathbb{C}^{L_k \times 1},
\end{equation}
where $L_k$ is the total number of transmitting paths of user $k$.

Similarly, for the $m$th FA on the RS of the FAR, we assume the elevation and azimuth angle of arrival (AoA) of the $q$th path are $\theta_{k,U}^q \in [0,\pi]$ and $\phi_{k,U}^q \in [0,\pi]$, respectively. Hence, the corresponding wave vector is $\bm{n}_{k,U}^q = [\mathrm{sin}\theta_{k,U}^q \mathrm{cos}\phi_{k,U}^q,\mathrm{cos}\theta_{k,U}^q]^T$. If we denote the subscript $U_m$ as the $m$th FA at the RS of FAR, the phase difference of the FA with the coordinate $\bm{r}_{U_m}$ can be given as $\rho_{k,U_m}^q ({\bm{r}}_{U_m})$. Then, the FRV of the $m$th FA to receive signals from user $k$ is
\begin{align}\label{FA receive field response vector}
    \bm{f}_{k,U} (\bm{r}_{U_m})  
    &= [\rho_{k,U_m}^1 ({\bm{r}}_{U_m}),\dots, \rho_{k,U_m}^{L_{U}} ({\bm{r}}_{U_m})]^T 
    \in \mathbb{C}^{L_{U} \times 1},
\end{align}
where $L_U$ is the total number of receiving paths of FAR.

Thus, the field response matrix (FRM) of all the $M$ FAs at RS of FAR can be given as
\begin{equation}
    \bm{F}_{k,U} (\bm{R}_{U})  
    = [\bm{f}_{k,U} (\bm{r}_{U_1}), \dots, \bm{f}_{k,U} (\bm{r}_{U_M})] 
    \in \mathbb{C}^{L_{U} \times M},
\end{equation}
where position matrix $\bm{R}_{U} = [\bm{r}_{U_1},\dots, \bm{r}_{U_M}] $ is the matrix representing the positions of all the $M$ FAs.

Furthermore, we define the path response matrix\cite{zhu2023modeling} from the reference point at the transmitting region to that of the receiving region as $\bm{\Sigma}_k \in \mathbb{C}^{L_{U} \times L_k} $, where $\bm{\Sigma}_{k_{p,q}}$ is the element in the $p$th row and $q$th column of $\bm{\Sigma}_k$, describing the path response between the $p$th transmitting path and the $q$th receiving path. 

As a result, the channel vector from the transmitter of user $k$ to the RS of FAR can be given as
\begin{equation}
    \bm{h}(\mathbf{t}_k, \bm{R}_{U}) = [\bm{F}_{k,U} (\bm{R}_{U})]^H 
 \bm{\Sigma}_k \bm{u}_k(\bm{t}_k) \in \mathbb{C}^{M \times 1}.
\end{equation}

We consider an amplify-and-forward FAR, the relay matrix is represented by a diagonal relay gain matrix $\bm{F}$\cite{behbahani2008optimizations}, where $\bm{F} = \mathrm{diag}(\mathop{f}_1,\dots,f_M) \in \mathbb{C}^{M \times M}$. Without loss of generality, we assume that $f_i = F, \forall i \in \mathcal{M}$.

After receiving the signals from the users, the $M$ FAs at the TS of FAR transmit the signals to the BS. Similar to \eqref{FA receive field response vector}, the FRV of the $m$th FA of the FAR to transmit the signal to the BS can be defined as
\begin{equation}
    \bm{f}_{B} (\bm{t}_{B_m})  
    = [\rho_{B_m}^1 ({\bm{t}}_{B_m}),\dots, \rho_{B_m}^{L_{B}} ({\bm{t}}_{B_m})]^T 
    \in \mathbb{C}^{L_{B} \times 1},
\end{equation} \label{FA transmit field response vector}where $L_{B}$ is the total number of transmitting paths of FAR, $\bm{t}_{B_m}$ is the position vector of the $m$th FA, and $\rho_{B_m}^p(\bm{t}_{B_m})$ is the phase difference of the $p$th transmitting path.

Thus, the FRM for the TS of FAR is written as
\begin{align}
    \bm{F}_{B} (\bm{T}_{B})  
    &= [\bm{f}_{B} (\bm{t}_{B_1}), \dots, \bm{f}_{B} (\bm{t}_{B_M})] \in \mathbb{C}^{L_{B} \times M},
\end{align}
where $\bm{T}_B = [\bm{t}_{B_1},\dots,\bm{t}_{B_N}]$.

Similarly, the FRV of the $n$th FA of the BS can be given as 
\begin{equation}\label{BS receive field response vector}
    \bm{b} (\bm{r}_{n})  
    = [\rho_{n}^1 ({\bm{r}}_{n}),\dots, \rho_{n}^{L_{b}} ({\bm{r}}_n)]^T \in \mathbb{C}^{L_{b} \times 1},
\end{equation}
where $L_{b}$ is the total number of the receiving paths at the BS, $\bm{r}_n$ is the two-dimensional coordinate vector of the $n$th FA of the BS, $\rho_n^q (\bm{r}_n)$ is the $q$th transmitting path of FAR.

Thus, the FRM of all the $N$ FAs at BS is written as
\begin{equation}
    \bm{B} (\bm{R})  
    = [\bm{b} (\bm{r}_{1}), \dots, \bm{b} (\bm{r}_{N})] \in \mathbb{C}^{L_{b} \times N},
\end{equation}
where $\bm{R} = [\bm{r}_1,\dots,\bm{r}_N]$ is the position matrix of the $N$ FAs.

As a result, the channel matrix from the FAR to the receiver at the BS is given by
\begin{equation}
    \bm{H}(\bm{T}_B, \bm{R}) = [\bm{B}(\bm{R})]^H \bm{\Sigma} \bm{F}_B (\bm{T}_B) \in \mathbb{C}^{N \times M},
\end{equation}
where $\bm{\Sigma}$ is the path response between FAR and BS.

As a result, the received signal in the BS can be given as,
\begin{align}
\nonumber
    \bm{y}_{BS} 
    &= \sum_{k=1}^K \underbrace{\bm{H}(\bm{T}_B, \bm{R}) \bm{F}}_{\triangleq\tilde{\bm{H}}} \ \underbrace{\bm{h}(\bm{t}_k,\bm{R}_{U})}_{\triangleq\tilde{\bm{h}}_k} p_k s_k \\
    &+\tilde{\bm{H}} \bm{\sigma}_U + \bm{\sigma}_B, 
\end{align}
where $p_k$ is the transmission power of user $k$, satisfying $p_k \leq P_{k}^{max}$, $s_k$ represents the information symbol for user $k$, which is modeled as $\mathcal{CN}(0,1)$, $\bm{\sigma}_U \sim \mathcal{CN}(\bm{0},\sigma_U^2 \bm{I}_M)$ is the additive white Gaussian noise (AWGN) at the RS of FAR, and $\bm{\sigma}_B \sim \mathcal{CN}(\bm{0},\sigma_B^2 \bm{I}_M)$ is the AWGN at the BS. 

\subsection{Problem Formulation}
The achievable rate of user $k$ can be given by $R_k =  \rm{log} (1 + \gamma_k)$, where $\gamma_k$ is the signal-to-interference-plus-noise-ratio (SINR) and can be given as
\begin{equation}\label{SINR}
    \gamma_k = \dfrac{||p_k \bm{\omega}_k^H \bm{\tilde{\bm{H}} \tilde{\bm{h}}}_k||_2^2}{(F\sigma_U^2+\sigma_B^2)||\bm{\omega}_k^H||_2^2 + \sum_{i=1,i\neq k}^{K}||p_i \bm{\omega}_k^H \bm{\tilde{\bm{H}} \tilde{\bm{h}}}_i||_2^2}, 
\end{equation}
where $\bm{\omega}_k$ is the receiving beamforming vector for user $k$. For convenience, we employ the equal-gain combining (EGC) method\cite{rizvi2010performance} and set a typical $\bm{\omega}_k = \bm{1}$ to obtain the received signal. As a result, the simplified SINR can be given as

\begin{equation}\label{SINR2}
    \gamma_k = \dfrac{||p_k \bm{\tilde{\bm{H}} \tilde{\bm{h}}}_k||_2^2}{\underbrace{F\sigma_U^2+\sigma_B^2}_{\triangleq \sigma^2} + \sum_{i=1,i\neq k}^{K}||p_i \bm{\tilde{\bm{H}} \tilde{\bm{h}}}_i||_2^2}, 
\end{equation}
which is influenced by $\bm{t}_k$, $\bm{R}_U$, $\bm{T}_B$, and $\bm{R}$.

Mathematically, the max-min fairness problem can be formulated as follows:
\begin{subequations}\label{sys2max0} 
\begin{align}
  \mathop{\max}_{ \bm{t}_k, \bm{R}_U, \bm{T}_B, \bm{R}} \;&  
   \mathop{\min}_{k \in \mathcal{K}} {\{R_k\}}\\
 \textrm{s.t.} \quad\:
   & \bm{t}_k \in \mathcal{C}_k\\
   & \bm{r}_{U_m} \in \mathcal{C}_U, \forall m \in \mathcal{M}, \\
   & \bm{t}_{B_m} \in \mathcal{C}_B, \forall m \in \mathcal{M},  \\
   &  \bm{r}_{n} \in \mathcal{C}_b, \forall n \in \mathcal{N},\\
   & ||\bm{r}_{U_m} - \bm{r}_{U_l}||_2 \geq d_0 , \forall m,l \in \mathcal{M} , m \neq l \\
   & ||\bm{t}_{B_m} - \bm{t}_{B_l}||_2 \geq d_0 , \forall m,l \in \mathcal{M} , m \neq l \\
   & ||\bm{r}_{n} - \bm{r}_{l}||_2 \geq d_0 , \forall n,l \in \mathcal{N} , n \neq l \\
   & 0\leq p_k\leq P_k, \quad \forall k \in \mathcal{K},
\end{align}
\end{subequations}
where $\mathcal{C}_k$, $\mathcal{C}_U$, $\mathcal{C}_B$, and $\mathcal{C}_b$ denote the given 2D regions, without loss of generality, all set as square regions with size $A \times A$, within which the FAs can switch the positions freely, respectively, and $d_0$ is the minimum required distance between FAs to avoid mutual coupling\cite{ye2023fluid}.
\begin{algorithm}[t]
\caption{Iterative algorithm for \eqref{sys2max2}}\label{Ialgorithm}

\KwIn{User number $K$, FA's positions $\bm{t}_1^{(0)}$ \dots $\bm{t}_K^{(0)}$, $\bm{R}_U^{(0)}$, $\bm{T}_B^{(0)}$, and $\bm{R}^{(0)}$, maximal transmitting power $P_1,\dots P_K$, desirable accuracy $\epsilon$, and iteration number $i = 0$.}
\KwOut{Optimal positions of FAs $\bm{t}_1^{I}$ \dots $\bm{t}_K^{I}$, $\bm{R}_U^{I}$, $\bm{T}_B^{I}$, and $\bm{R}^{I}$, and maximal $\alpha^I$.}
Set $j = 1$. 

\While{$j  \leq K$}{Calculate effective channel gain $g_i = ||p_j \bm{\tilde{\bm{H}} \tilde{\bm{h}}}_j||_2^2$.

$j = j + 1$.}

$i = i+1$, $\alpha^{(i)} = \mathrm{min}\{g_k\}_{k=1}^K$.

\While{$\alpha^{(i)} - \alpha^{(i-1)} \geq \epsilon$}{

Update the position parameters by using Algorithm~\ref{Aalgorithm} with given $\bm{t}_1^{(i)}$, \dots, $\bm{t}_K^{(i)}$, $\bm{R}_U^{(i)}$, $\bm{T}_B^{(i)}$, and $\bm{R}^{(i)}$. Set $j = 1$. 

\While{$j  \leq K$}{Calculate effective channel gain $g_i = ||p_j \bm{\tilde{\bm{H}} \tilde{\bm{h}}}_j||_2^2$.

$j = j + 1$.}

$I = i-1$, $i = i+1$, $\alpha^{(i)} = \mathrm{min}\{g_k\}_{k=1}^K$.
}
\end{algorithm}
\section{Algorithm Design}
Given that the achievable rate is related to its corresponding SINR, we first try to determine the worst SINR, as given in \eqref{SINRCompare}. Since $\sigma^2$ and $||p_i \bm{\tilde{\bm{H}} \tilde{\bm{h}}}_i||_2^2$ are all positive, we can obtain the following equivalence relationship,
\begin{figure*}
    \begin{align}\label{SINRCompare}
    \noindent
    \gamma_m - \gamma_n    
    &= \dfrac{\left(\sigma^2+\sum_{i=1}^K||p_i \bm{\tilde{\bm{H}} \tilde{\bm{h}}}_i||_2^2\right)\left(||p_m \bm{\tilde{\bm{H}} \tilde{\bm{h}}}_m||_2^2 - ||p_n \bm{\tilde{\bm{H}} \tilde{\bm{h}}}_n||_2^2 \right)}{\left(\sigma^2 + \sum_{i=1,i\neq m}^{K}||p_i \bm{\tilde{\bm{H}} \tilde{\bm{h}}}_i||_2^2\right)\left(\sigma^2 + \sum_{i=1,i\neq n}^{K}||p_i \bm{\tilde{\bm{H}} \tilde{\bm{h}}}_i||_2^2\right)}, \forall m,n \in \mathcal{K}, m \neq n.
    \end{align}
%{\noindent} \rule[-10pt]{19cm}{0.05em}
\hrulefill

\end{figure*}

\begin{equation}\label{newOBJ}
  \mathop{\min}_{k \in \mathcal{K}} {\{R_k\}} \iff  \mathop{\min}_{k \in \mathcal{K}} {\{\gamma_k\}} \iff \mathop{\min}_{k \in \mathcal{K}} \{||p_k \bm{\tilde{\bm{H}} \tilde{\bm{h}}}_k||_2^2\}.
\end{equation}

Due to the fact that the power control problem can be solved via the conventional method such as interference control method, it is not considered in this paper. Then, the difficulties to solving this include the objective function is still non-convex and the constraints (\ref{sys2max0}{f}) - (\ref{sys2max0}{h}) are all non-convex either. 

To deal with the non-convex objective function \eqref{newOBJ}, an auxiliary variable $\alpha$ is introduced to convert the original max-min problem to the following problem equivalently,
\begin{subequations}\label{sys2max2} 
\begin{align}
  \mathop{\max}_{ \bm{t}_k, \bm{R}_U, \bm{T}_B, \bm{R}} \;&  \alpha, \\
 \textrm{s.t.} \quad\:
   & (\ref{sys2max0}\rm{b}) - (\ref{sys2max0}\rm{h})\\
   & {||p_k \bm{\tilde{\bm{H}} \tilde{\bm{h}}}_k||_2^2} \geq \alpha, \forall  k \in \mathcal{K}.
\end{align}
\end{subequations}

To solve this problem, we propose an iterative algorithm as shown in Algorithm~\ref{Ialgorithm}.

When Algorithm~\ref{Ialgorithm} determines the user $k$ owning the worst effective channel gain in certain iteration, we propose the following optimization problem to maximize it as
\begin{subequations}\label{sys2max3} 
\begin{align}
  \mathop{\max}_{ \bm{t}_k, \bm{R}_U, \bm{T}_B, \bm{R}} \;&  ||\bm{\tilde{\bm{H}} \tilde{\bm{h}}}_k||_2^2, \\
 \textrm{s.t.} \quad\:
   & (\ref{sys2max0}\rm{b}) - (\ref{sys2max0}\rm{h})\\
   & {||p_i \bm{\tilde{\bm{H}} \tilde{\bm{h}}}_i||_2^2} \geq \alpha_0, \forall  i \in \mathcal{K}, i \neq k.
\end{align}
\end{subequations} 

Subsequently, we provide an alternation optimization method, as shown in Algorithm~\ref{Aalgorithm} to tackle problem \eqref{sys2max3} by iteratively address sub-problems, where each sub-problem optimizes only one variable while keeping the others constant.

\subsection{Optimization on the FA's position of user $k$}
Given that the change of $\bm{t}_k$ only changes the channel gain of user $k$, the sub-problem for optimization of $\bm{t}_k$ is given by
\begin{subequations}\label{sub1} 
\begin{align}
  \mathop{\max}_{ \bm{t}_k} \;&  \mathrm{tr}(\bm{u}_k^H (\bm{t}_k) \bm{\Upsilon}_k \bm{u}_k (\bm{t}_k)), \\
 \textrm{s.t.} \quad\:
   & (\ref{sys2max0}\rm{b}) ,
\end{align}
\end{subequations}
where $\bm{\Upsilon}_k = \bm{\Sigma}_k^H \bm{F}_{k,U} \tilde{\bm{H}}^H \tilde{\bm{H}} \bm{F}_{k,U}^H \bm{\Sigma}_k $. Problem \eqref{sub1} is not concave because of the non-concave objective function. Considering the fact that \eqref{sub1} is convex with respect to $\bm{u}_k(\bm{t}_k)$, we give the lower bound of (\ref{sub1}a) by deriving the first-order Taylor expansion of given point $\bm{t}_k^{(i)}$ \cite{wu2018joint}, where the subscript $(i)$ means the $i$th iteration of SCA,

\begin{align}\label{OBJsub11}
\nonumber
&(\ref{sub1}a) 
 \geq\bm{u}_k^H (\bm{t}_k^{(i)}) \bm{\Upsilon}_k \bm{u}_k (\bm{t}_k^{(i)}) \\
 \nonumber
 &+ 2Re\left\{\bm{u}_k^H (\bm{t}_k^{(i)}) \bm{\Upsilon}_k \left(\bm{u}_k (\bm{t}_k)-\bm{u}_k(\bm{t}_k^{(i)} )\right)\right\} \\
&= \underbrace{2Re\left\{\bm{u}_k^H (\bm{t}_k^{(i)}) \bm{\Upsilon}_k \bm{u}_k (\bm{t}_k)\right\}}_{\triangleq \upsilon(\bm{t}_k)} - \underbrace{\bm{u}_k^H (\bm{t}_k^{(i)}) \bm{\Upsilon}_k \bm{u}_k (\bm{t}_k^{(i)})}_{constant}.
\end{align}

On this basis, maximizing (\ref{sub1}a) can be transformed into maximizing \eqref{OBJsub11}. Unfortunately, $\upsilon(\bm{t}_k)$ is still either convex nor concave over $\bm{t}_k$. Based on \eqref{OBJsub11}, we can derive a quadratic surrogate function to globally lower-bound $\upsilon(\bm{t}_k)$ as
\begin{align}\label{OBJsub111}
    \nonumber
&\upsilon(\bm{t}_k^{(i)})  + \nabla (\upsilon(\bm{t}_k^{(i)}))^T (\bm{t}_k - \bm{t}_k^{(i)}) - \frac{\delta_k}{2} (\bm{t}_k - \bm{t}_k^{(i)})^T (\bm{t}_k - \bm{t}_k^{(i)})\\ 
    \nonumber
    &= \underbrace{- \frac{\delta_k}{2} \bm{t}_k^T \bm{t}_k  + (\nabla \upsilon(\bm{t}_k^{(i)}) + \delta_k \bm{t}_k^{(i)})^T \bm{t}_k}_{\triangleq \hat{\upsilon}(\bm{t}_k)} \\
    &+ \underbrace{\upsilon(\bm{t}_k^{(i)}) - \frac{\delta_k}{2} (\bm{t}_k^{(i)})^T (\bm{t}_k^{(i)})}_{constant},
\end{align}
where $\nabla$ is the gradient vector and $\delta_k$ is a positive number making $\delta_k \bm{I}_2 \succeq \nabla^2 \upsilon(\bm{t}_k)$ with its closed form given in \cite{ma2023mimo}. Substituting (\ref{sub1}a) by \eqref{OBJsub111}, this sub-problem is concave and can be solved by existing tool box.

\begin{algorithm}[t]
\caption{Alternating optimization algorithm for \eqref{sys2max3}}\label{Aalgorithm}

\KwIn{FA's positions and maximal transmitting power when Algorithm~\ref{Ialgorithm} invokes this algorithm, maximal iteration number $I_{max}$, iteration number $i = 0$, and $\alpha_0$.} 
\KwOut{Position of FAs $\bm{t}_1^{i}$ \dots $\bm{t}_K^{i}$, $\bm{R}_U^{i}$, $\bm{T}_B^{i}$, and $\bm{R}^{i}$.}

\While{$i \leq I_{max}$}{
Solve problem \eqref{sys2max3} through solving a set of convex forms of sub-problems \eqref{sub1} \eqref{sub2} \eqref{sub3} \eqref{sub4} using the SCA method.

\If{\eqref{sys2max3} converges}{Break the loop.}
$i = i+1$.
}
\end{algorithm}

\subsection{Optimization on the $m$th FA at the RS of FAR}
When given $\bm{t}_k$, $\bm{T}_{B}$, $\bm{R}$ and $\{\bm{r}_{U_i},i\neq m\}_{i=1}^M$, the sub-problem for optimization of $\bm{r}_{U_m}$ is given by
\begin{subequations}\label{sub2} 
\begin{align}
  \mathop{\max}_{ \bm{r}_{U_m} }\;&   \mathrm{tr}\left(\bm{F}_{k,U}(\bm{r}_{U_m}) \bm{\lambda}_k \bm{\lambda}_k^H \bm{F}_{k,U}^H(\bm{r}_{U_m}) \bm{\Lambda} \right) , \\
 \textrm{s.t.} \quad\:
   & (\ref{sys2max0}\rm{c}), (\ref{sys2max0}\rm{f}), (\ref{sys2max3}\rm{c}),
\end{align}
\end{subequations}
where $\bm{\lambda}_k = \bm{\Sigma}_k \bm{u}_k$, and $\bm{\Lambda} = \tilde{\bm{H}}^H \tilde{\bm{H}}$. Furthermore, if denote $\lambda_k(i)$ is the $i$th element of $\bm{\lambda}_k$, the objective function can be equivalently given by \eqref{sub2OBJ1}.
\begin{figure*}
    \begin{align}\label{sub2OBJ1}
    (\ref{sub2}a)    
    \nonumber 
    &= {\underbrace{\mathrm{tr}\left[\lambda(m)\lambda^H(m) \bm{f}_{k,U}(\bm{r}_{U_m})  \bm{f}^H_{k,U}(\bm{r}_{U_m}) \bm{\Lambda} \right]}_{\triangleq {g}(\bm{r}_m)}}  + \mathrm{tr}\left[\lambda(m)\bm{f}_{k,U}(\bm{r}_{U_m}) \sum_{i=1,i\neq m}^M\lambda^H(i) \bm{f}^H_{k,U}(\bm{r}_{U_i}) \bm{\Lambda} \right]\\ 
    &+ \mathrm{tr}\left[\bm{f}^H_{k,U}(\bm{r}_{U_m}) \bm{\Lambda}  \underbrace{\sum_{i=1,i\neq m}^M \lambda(i)\bm{f}_{k,U}(\bm{r}_{U_i}) \lambda(m)^H}_{\triangleq \bm{g}_1} \right] + \underbrace{\mathrm{tr}\left[ \sum_{i=1,i\neq m}^M \lambda(i)\bm{f}_{k,U}(\bm{r}_{U_i}) \sum_{j=1,i\neq m}^M \lambda(j)^H \bm{f}^H_{k,U}(\bm{r}_{U_j}) \bm{\Lambda} \right]}_{\triangleq {g_2}}.
    \end{align}
%{\noindent} \rule[-10pt]{19cm}{0.05em}
%\hrulefill

\end{figure*}
It can be found that \eqref{sub2OBJ1} is convex over $\bm{f}(\bm{r}_{U_m})$, hence we can utilize first-order Taylor expansion with the given point $\bm{r}^{(i)}_{U_m}$, to approximate $g(\bm{r}_{U_m})$ with a lower bound $\hat{g}(\bm{r}_{U_m})$ as shown in \eqref{sub2OBJ22}. $\hat{g}(\bm{r}_{U_m})$ is convex with respect to $\bm{f}_{k,U}(\bm{r}_{U_m})$ but neither convex nor concave over $\bm{r}_{U_m}$. Fortunately, it has a close form to \eqref{OBJsub11}, hence we can get a surrogate function by employing the second-order Taylor expansion to obtain its lower bound the same method as \eqref{OBJsub111} employs.
\begin{figure*}
    \begin{align}\label{sub2OBJ22}
    {g}(\bm{r}_{U_m})    
    \nonumber 
    &\geq  g(\bm{r}_{U_m}^{(i)}) + 2\mathrm{Re}\left\{\lambda(m)\lambda^H(m) \left(\bm{f}^H_{k,U}(\bm{r}_{U_m}) - \bm{f}^H_{k,U} (\bm{r}_{U_m}^{(i)})\right) \bm{\Lambda}  \bm{f}_{k,U} (\bm{r}_{U_m}^{(i)}) + \bm{f}^H_{k,U}(\bm{r}_{U_m})  \bm{\Lambda} \bm{g}_1\right\} + g_2 \\
    &= \underbrace{2\mathrm{Re}\left\{\bm{f}^H_{k,U}(\bm{r}_{U_m}) \left(\lambda(m)\lambda^H(m)  \bm{\Lambda} \bm{f}_{k,U} (\bm{r}_{U_m}^{(i)}) +  \bm{\Lambda} \bm{g}_1 \right)  \right\}}_{\triangleq \hat{g}(\bm{r}_{U_m})}\underbrace{ - g(\bm{r}_{U_m}^{(i)}) + g_2}_{ constant}.
    \end{align}
%{\noindent} \rule[-10pt]{19cm}{0.05em}
\hrulefill

\end{figure*}
For the non-convex constraints, (\ref{sys2max3}c) has the similar form to (\ref{sys2max3}a), hence it can be transformed into convex, and (\ref{sys2max0}f) can be approximated by maximizing its first-order Taylor expansion at the given point $\bm{r}_{U_m}^{(i)}$ as
\begin{equation}\label{constraint}
 ||\bm{r}_{U_m} - \bm{r}_{U_l}||_2  \geq  \frac{(\bm{r}_{U_m}^{(i)} - \bm{r}_{U_l})^T(\bm{r}_{U_m} - \bm{r}_{U_l})}{||\bm{r}_{U_m}^{(i)} - \bm{r}_{U_l}||_2}.
\end{equation}

Then the sub-problem can be solved by existing tool box.
\subsection{Optimization on the $m$th FA at the TS of FAR and the $n$th FA at BS}
In this subsection, we will prove that the optimization of the $m$th FA's position at FAR with given $\bm{t}_k$, $\bm{R}_{U}$, $\bm{R}$ and $\{\bm{t}_{B_i},i\neq m\}_{i=1}^M$ and the optimization of the $n$th FA's position at BS with given $\bm{t}_k$, $\bm{R}_{U}$, $\bm{T}_{B}$ and $\{\bm{r}_{i},i\neq n\}_{i=1}^N$ both can be formulated with the similar form to \eqref{sub2}, therefore can be addressed with the same approach.

For the optimization of the $m$th FA at TS of FAR, the sub-problem can be formulated as
\begin{subequations}\label{sub3} 
\begin{align}
  \mathop{\max}_{ \bm{t}_{B_m} }\;&   \mathrm{tr}\left(\tilde{\bm{h}}_{k}^H \bm{F}^H_{B}(\bm{t}_{B_m}) \bm{\Phi} \bm{F}_{B}(\bm{t}_{B_m}) \tilde{\bm{h}}_{k} \right) , \\
 \textrm{s.t.} \quad\:
   & (\ref{sys2max0}\rm{d}), (\ref{sys2max0}\rm{g}), (\ref{sys2max3}\rm{c}),
\end{align}
\end{subequations}
where $\bm{\Phi} = \bm{\Sigma}^H \bm{B}(\bm{R}) \bm{B}^H(\bm{R}) \bm{\Sigma}$. 

For the optimization of the $n$th FA at BS, the sub-problem can be formulated as
\begin{subequations}\label{sub4} 
\begin{align}
  \mathop{\max}_{ \bm{r}_{n} }\;&   \mathrm{tr}\left(\bm{\phi}_{k}^H \bm{B}(\bm{r}_n) \bm{B}^H(\bm{r}_n) \bm{\phi}_{k} \right) , \\
 \textrm{s.t.} \quad\:
   & (\ref{sys2max0}\rm{e}), (\ref{sys2max0}\rm{h}), (\ref{sys2max3}\rm{c}),
\end{align}
\end{subequations}
where $\bm{\phi}_k = \bm{\Sigma} \bm{F}_B(\bm{T}_B) \tilde{\bm{h}}_k$. 
\addtolength{\topmargin}{-0.02in}

According to the equalities of
\begin{equation}
(\ref{sub3}a) = \mathrm{tr}\left( \bm{F}_{B}(\bm{t}_{B_m}) \tilde{\bm{h}}_{k} \tilde{\bm{h}}_{k}^H \bm{F}^H_{B}(\bm{t}_{B_m}) \Phi \right),    
\end{equation}
\begin{equation}
(\ref{sub4}a) = \mathrm{tr}\left( \bm{B}(\bm{r}_n)  \bm{\phi}_{k} \bm{\phi}_{k}^H \bm{B}^H(\bm{r}_n) \right),    
\end{equation}
both (\ref{sub3}a) and (\ref{sub4}a) keep the same form to (\ref{sub2}a), and they can be approximated by their respective second-order Taylor expansion which are convex with respect to $\bm{t}_{B_m}$ and $\bm{r}_n$, respectively. The non-convexity of (\ref{sys2max0}\rm{g}) and (\ref{sys2max0}\rm{h}) can be addressed with their corresponding first-order Taylor expansions.

\subsection{Complexity Analysis}
%Without loss of generality, we assume the number of paths (i.e., $L_k$, $L_U$, $L_B$, and $L_b$) and the number of FAs ($M$ and $N$) are of the same order of magnitude. 
The complexity of calculating Algorithm \ref{Aalgorithm} is $\mathcal{O}(ML_U^2\gamma_1 + M^{2.5}\mathrm{ln}(\frac{1}{\varepsilon})\gamma_2)$\cite{ma2023mimo}, where $\varepsilon$ is the accuracy for the interior-point
method, $\gamma_1$ is the maximum number of
inner iterations of solving each sub-problem, and $\gamma_2$ is the maximum number of iterations required to tackle the quadratic
programming (QP) problem. Hence the complexity of solving Algorithm~\ref{Ialgorithm} is $\mathcal{O}(NM^2I + IML_U^2\gamma_1 + IM^{2.5}\mathrm{ln}(\frac{1}{\varepsilon})\gamma_2)$, where $I$ is the number of iterations of Algorithm~\ref{Ialgorithm}.

\section{Simulation Results}
In our simulations, we assume all the elevation angles and azimuth angles are independent and identically distributed (i.i.d), randomly distributed in $[0,\pi]$, and the path matrix is diagonal with $\bm{\Sigma}_{k_{1,1}} \sim \mathcal{CN}(0,\beta/(\beta + 1))$ and $\bm{\Sigma}_{k_{l,l}} \sim \mathcal{CN}(0,1/(\beta + 1)(L-1)), \forall l = 2,\dots,L$, where $\beta = 1$\cite{ding2015performance} represents the ratio of the average power of the line-of-sight (LoS) path to the average power of the non-line-of-sight (NLoS) path and $L = L_k = L_U = L_B = L_b = 4$ is the total number of transmitting/receiving paths. For the parameters related to FAs, we set $M = 4$, $N = 5$, $d_0 = \lambda/2$ and $A = 4\lambda$. Assume all the users contains the same maximum transmitting power $P^{max}$, and the average SNR is defined as $||P^{max}||_2^2/\sigma^2 = 5$ $\rm{dB}$.

\begin{figure}[t]
    \centering    \includegraphics[width=1\linewidth]{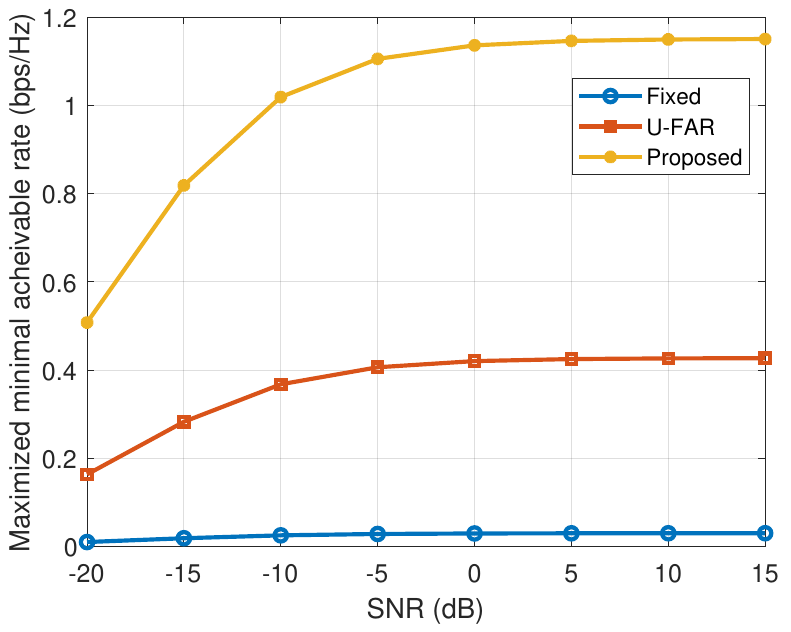}
    \caption{Maximized minimal rate vers. SNR}
    \label{SimulationSNR}
\end{figure}
In the two legends, ``Fixed'' represents the achievable rate of the user with the worst communication conditions when all antennas have fixed positions. ``U-FAR'' denotes the situation in which only FAs of FAR can adjust their positions, which partly optimizes the minimal user rate of the system. ``Proposed'' denotes our proposed iterative alternating algorithm.

In Fig~\ref{SimulationSNR}, we compares the maximized minimal rate with respect to average SNR. We can observe that for all three baselines, as the SNR increases, all schemes have a higher maximized minimal rate, and the proposed method keeps the best maximized minimal rate over all kinds of SNR.  Furthermore, ``Fixed'' scheme always holds a minimal rate close to zero indicating the user with minimal rate indeed endures rather harsh communication conditions, and the position optimization of FAs indeed improves the performance. 

\begin{figure}[t]
    \centering
    \includegraphics[width=1\linewidth]{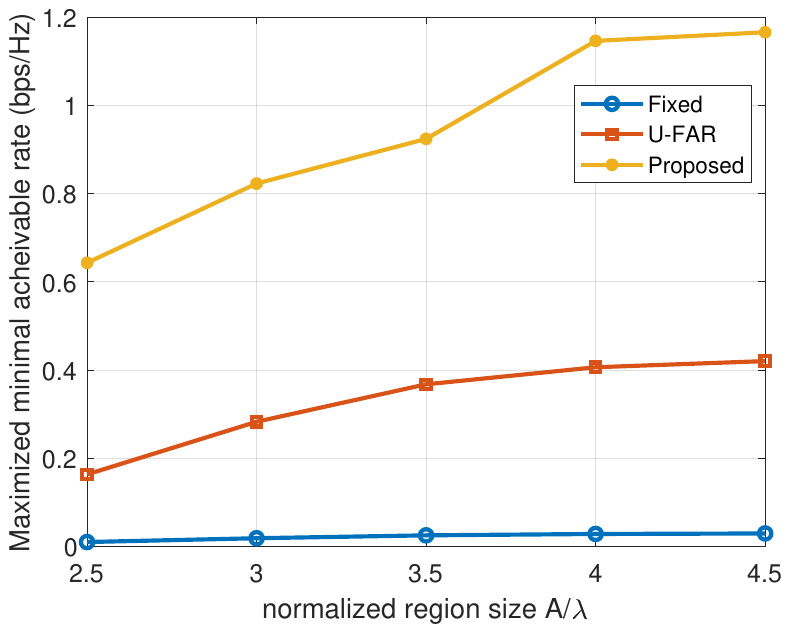}
    \caption{Maximized minimal rate vers. normalized region}
    \label{SimulationNormalizedregion}
\end{figure}
Fig~\ref{SimulationNormalizedregion} illustrates  the maximized minimal achievable rate over normalized region size, defined as the region size $A$ normalized by the wave length $\lambda$. As the normalized region size increases, the maximized minimum rate for all schemes improves, with the proposed method consistently achieving the highest maximized minimum rate across all region sizes. This indicates that optimizing antenna positions can significantly enhance system performance, especially in larger regions where the benefits of optimization are more pronounced.

\section{Conclusion}
In this paper, we investigate the uplink multi-user MISO communication with FAR assisted. To ensure the rate of weak user in the system, we formulate the max-min fairness problem. We first transform this problem into a maximization problem and propose an alternating algorithm to iteratively address it. Simulations results demonstrate the proposed method can notably improve the achievable rate of the weak user over given SNR range and normalized region size range.

\bibliographystyle{IEEEtran}
\bibliography{IEEEabrv,FAR}
\end{document}